\documentclass[prl,aps,twocolumn,showpacs,groupedaddress,superscriptaddress]{revtex4}

\usepackage{graphicx,epsf,bm,bbm,color,amsmath,ulem,amssymb}

  \newcommand{\tr}[1]{\textrm{#1}}
	\newcommand{\h}[1]{\hat{#1}}

	\newcommand{\beq}{\begin{eqnarray}}
	
	\newcommand{\eeq}{\end{eqnarray}}
	\newcommand{\nn}{\nonumber}

\newcommand{\bem}{\begin{pmatrix}}
\newcommand{\eem}{\end{pmatrix}}
\newcommand{\hb}{\hbar}

\begin{document}

\title{Quantum Charge Pumps with Topological Phases in Creutz Ladder} 
 \author{Ning Sun and Lih-King Lim}
 \affiliation{Institute for Advanced Study, Tsinghua University, Beijing 100084, P. R. China}
\bigskip

\bigskip

\date{\today}

\begin{abstract}
Quantum charge pumping phenomenon connects band topology through the dynamics of a one-dimensional quantum system. In terms of a microscopic model, the Su-Schrieffer-Heeger/Rice-Mele quantum pump continues to serve as a fruitful starting point for many considerations of topological physics. Here we present a generalized Creutz scheme as a distinct two-band quantum pump model. By noting that it undergoes two kinds of topological band transitions accompanying with a Zak-phase-difference of $\pi$ and $2\pi$, respectively, various charge pumping schemes are studied by applying an elaborate Peierl's phase substitution. Translating into real space, the transportation of quantized charges is a result of cooperative quantum interference effect. In particular, an all-flux quantum pump emerges which operates with time-varying fluxes only and transports two charge units. This puts cold atoms with artificial gauge fields as an unique system where this kind of phenomena can be realized.
\end{abstract}
\maketitle

{\it Introduction\,-- } 
Quantum charge pumping was one of the early example of a counterintuitive one-dimensional (1D) transport phenomenon as a result of the subtle interplay between nontrivial band topology and quantum adiabatic transport \cite{Thouless83}.
It underpins the many facets of understanding topological physics ranging from the integer quantum Hall effect \cite{IQHE,TKKN82} to modern studies of electric polarization \cite{King93,Resta94}, and most recently, topological Floquet physics \cite{Yao07,Oka09,Kita10,Jiang11,Gu11,Lindner11,Hauke12,Delplace13,Rudner13,Zheng14,Kundu14}. Explicit models of a quantum pump, however, are surprisingly rare \cite{Xiao10, Shindou05, Fu06, Kraus12, Keselman13}. The prototype of its realization is the one based on the Su-Schrieffer-Heeger/Rice-Mele (SSH/RM) model \cite{SSH79, RM82} for polyacetylene. 
It has generated renewed interests thanks to various recent experimental realizations in ultracold atoms and condensed matter systems, where the Zak phase \cite{Atala2013}, quantum pumping phenomena \cite{Lohse16, Nakajima16, Lu16}, and chiral solitonic edge states \cite{Cheon15} are directly measured. 
In this Letter, we present a generalized Creutz scheme \cite{Creutz99} as a distinct microscopic two-band quantum pump model 
utilizing quantum interference effect. The model points to a new quasi-one-dimensional tight-binding quantum pump displaying rich topological physics, and realizable with artificial gauge fields in a cold atomic setting.

In the SSH/RM quantum pump, the physical picture is one based on confining quantum particle in a periodic potential that `slides' slowly in time \cite{Xiao10, Wang13}. Specifically, the sliding potential interpolates the two possible alternations of bond strength, such that two dimerization phases of the SSH are realized in a time periodic manner. 

\begin{figure}
\begin{center}
\includegraphics[width=8.8cm]{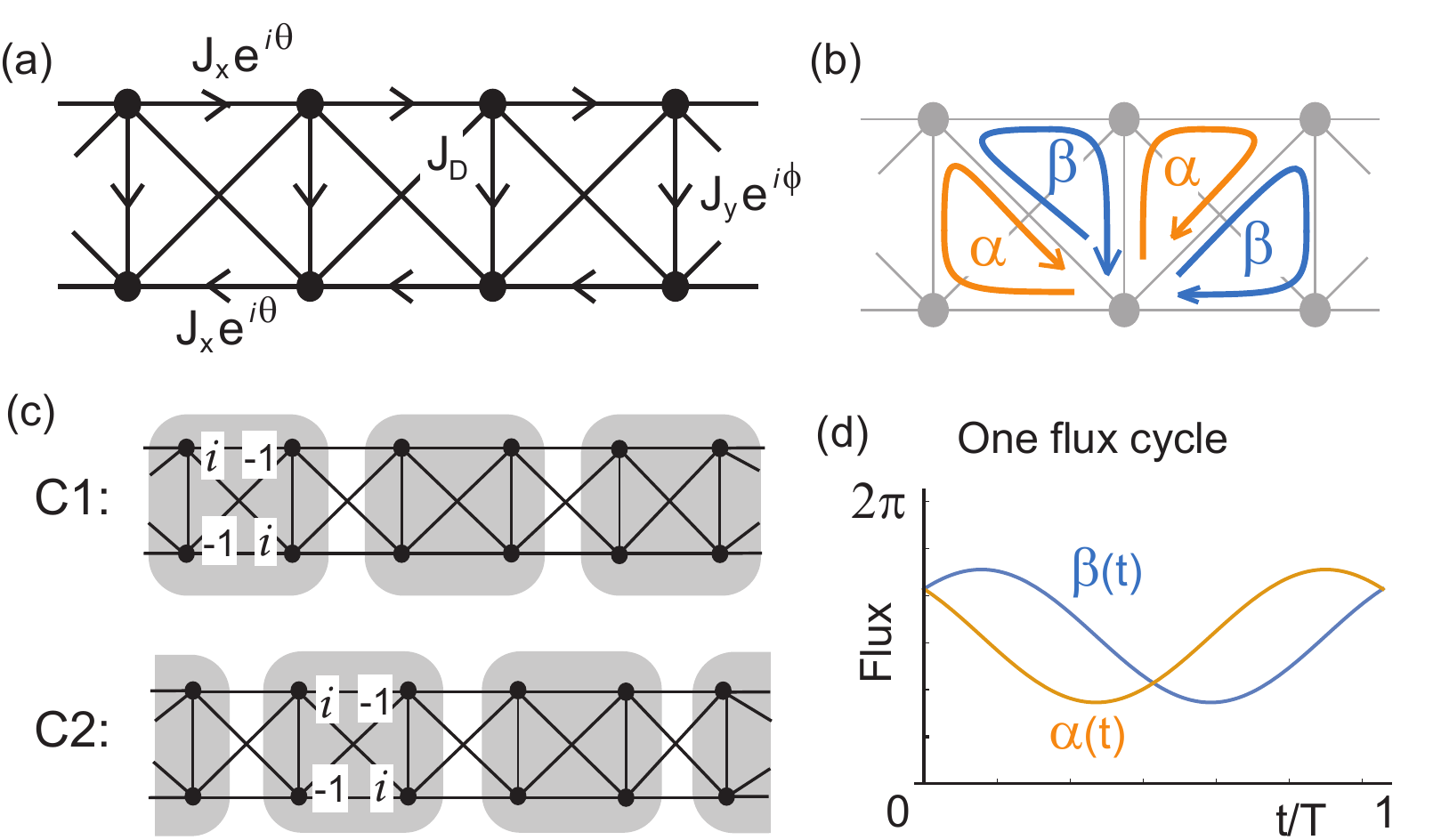}
\end{center}
\caption{(a) Generalized Creutz scheme with cross-link hopping $J_D$ and phase imprints $\theta$ and $\phi$ in a two-leg ladder lattice. (b) Physical flux pattern given by two gauge-invariant fluxes $\alpha=\theta-\phi$ and $\beta=\theta+\phi$. (c) Representatives of two plaquette coverings of the lattice in the topological regime. The values indicated are components of the (unrenormalized) wavefunction. (d) A typical flux cycle for an all-flux quantum pump.}
\end{figure}

In the quasi-1D Creutz model (Fig. 1a with $\phi=0$) \cite{Creutz99, Creutz01}, there actually exists  \textit{three} distinct phases, a feature not apparent by inspection of the tight-binding model. An intuition is provided as follows. First, notice that cross-link hoppings and external flux in the two-leg ladder lattice enhance quantum interference effect.
As a result, a complete basis in the topological regime, as found by Creutz \cite{Creutz99}, takes the form of plaquette state which spans two unit cells. This renders a full non-overlap covering of the lattice in two possible ways (denoted as C1, C2, Fig.1c). 
These, together with a trivial phase when the cross-linking is weak, form the basic topological features of the Creutz model \cite{Bermudez09,Hugel14,Sticlet14}.

Turning to the associated quantum pump, a new element is the introduction of an additional phase $\phi$ in the hopping along the rung of the ladder (Fig. 1a). In combination with the cross-linking, it shows an instance of Peierl's phase substitution with no simple external magnetic field representation (Fig. 1b).
The phase imprint $\phi$ opens the door in the Creutz model for various `adiabatic' connecting paths among the distinct phases \cite{Teo10,Roy11,Lopes16}, which are separated \textit{a priori} by band crossing, thereby allowing the realization of various quantum pumps. 
While some give pumping scenarios akin to the SSH/RM, remarkably, a novel kind emerges in the parameter plane of the two gauge-invariant fluxes wherein all bond strength are held \textit{fixed} - an all-flux quantum pump (Fig. 1d). It is associated with an accumulated Zak-phase-difference of $4\pi$ and a topological charge pumping of two units. This in turn can be understood as a connection between the two plaquette states coverings, a distinctive quantum interference feature of the Creutz model.
 
We will begin by clarifying the topological phases in the Creutz model with nontrivial flux patterns, emphasizing band crossings, rather than gap closing, which separates them. A generalized scheme opens the path for studying various quantum pumps with the associated phases. Then we discuss a realization scheme in cold atoms with time-dependent artificial gauge fields \cite{Dalibard11,Goldman14}.

{\it Topological characteristics of the Hamiltonian\,-- }
The generalized Creutz scheme is described by 
\beq
\h{H}&=&-\sum_{j} [J_X e^{i\theta} \h{a}_{j+1}^\dag \h{a}_{j}+J_X e^{-i\theta} \h{b}_{j+1}^\dag \h{b}_{j} + J_Y e^{-i\phi} \h{a}_j^\dag \h{b}_j 
\nn\\&&+ J_D \h{a}_j^\dag \h{b}_{j+1}+J_D \h{b}_j^\dag \h{a}_{j+1} +\tr{H.c.}]
\eeq                 
where $\h{a}_j^\dag (\h{a}_j)$ and $\h{b}_j^\dag (\h{b}_j)$ are creation (annihilation) operators on site $j$, belonging to the upper and lower chains, respectively. Compared with the usual ladder model, the additional ingredients are the cross-link diagonal hopping $J_D$ and the phase imprints $\theta$, $\phi$. 

The Bloch Hamiltonian is given by
\beq
H(k)=-\bem 2 J_X \cos(ka-\theta)& 2 J_D\cos(ka)+J_Y e^{-i\phi}\\
2 J_D\cos(ka)+J_Y e^{i\phi}& 2 J_X \cos(ka+\theta)\eem
\eeq
where $a$ is the lattice constant, and for short, we introduce the corresponding pseudospin components $H(k)=h_0(k)\,I+\bm{h}(k)\cdot\bm{\sigma}$, where $\{I,\bm{\sigma}\}$ are the identity and the three Pauli matrices, acting on the upper/lower chain space. It is a two-band model with the energy spectrum $E_{\pm}(k; J_{X,Y,D},\theta,\phi)=h_0(k)\pm|\bm{h}(k)|$ \cite{foot1}. 

The presence of the cross-link hopping alters the connectivity of the lattice in an interesting way. In comparison with a regular two-leg ladder $(J_D=0)$, the physical flux pattern is fixed by \textit{four} distinct shortest closed paths in the lattice (rather than one when $J_D=0$) and \textit{two} gauge-invariant fluxes $\alpha=\theta-\phi$ and $\beta=\theta+\phi$ (rather than simply $2\theta$ when $J_D=0$) with a physical range $-\pi<\alpha,\beta\leq\pi$, see Fig. 1b. In the convention chosen, therefore, the phase imprints $\theta$ and $\phi$ take range in the shaded area in Fig. 2b, the equivalent of the `first Brillouin zone'. As a side remark, the elaborate Peierl's phase substitution in the cross-link two-leg lattice results in inequivalent physical flux patterns overlapping in real space. It therefore cannot be related to a simple picture of an externally applied magnetic field or a magnetic background.  We now consider the bulk band topology for the original $(\phi=0)$ and the generalized $(\phi\neq 0)$ Creutz schemes with fixed $J_{X,Y}$. 
\begin{figure}
\begin{center}
\includegraphics[width=8.5cm]{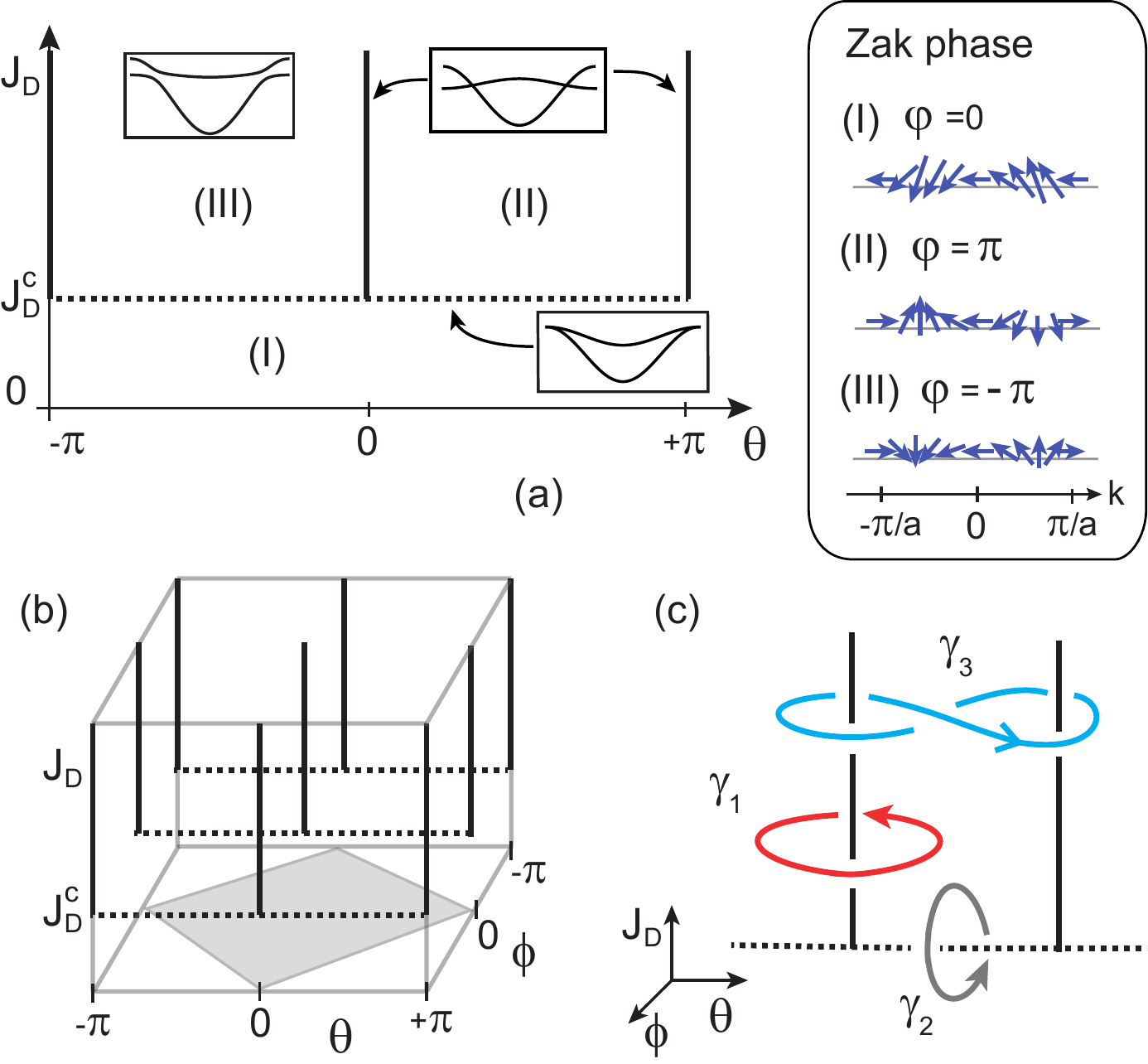}
\end{center}
\caption{(a) Left panel: Three phases of the Creutz model with finite $(J_X,J_Y)$ and $\phi=0$. Insets illustrate the generic two energy bands with or without band crossings. The boundaries (solid and dotted lines) are determined by the number of band crossings (two and one, respectively).  Right panel: Lower band eigenstates (arrows) showing different winding properties, described by the Zak phase. (b) The phase diagram in the extended $(\theta,\phi,J_D)$-parameter space. Shaded region is the physical range for $\theta$, $\phi$. (c) Various pumping schemes encircling the line degeneracies.}
\end{figure}

\textit{Case $\phi=0$:} In the $(\theta,J_D)$-plane we identify three regions, labeled as (I), (II) and (III), as being characterized generically by two energy bands with no \textit{band crossing} for all quasimomentum (the system need not be gapped at half filling), see Fig. 2a. The boundaries are given by a critical cross-link hopping $J_D^c=J_Y/2$ and the phase value $\theta=0,\pi$. 

What is not apparent from the energy bands are the three `phases' distinguished by their bulk band topology \cite{foot5} as characterized by the Zak phase $\varphi_{Zak}\equiv i\int_{1BZ} dk \,\langle u_k | \partial_k u_k\rangle$, i.e., the \textit{winding} of the lower band 2-spinor eigenstate $|u_k\rangle$ of $H(k)$ on the Bloch sphere as the quasimomentum parameter varies from $k=-\pi/a$ to $\pi/a$ in the first Brillouin zone $(1BZ)$ \cite{Zak89,foot2}. They are given by $\varphi_{Zak}^{\tr{I,II,III}}=0,\pi,-\pi$, respectively, see Fig. 2a \cite{Bermudez09,Hugel14,Junemann16}. 

A nontrivial Zak phase, in (II) and (III), signifies the existence of edge states in a finite-sized system \cite{Creutz99}. The reversal in the winding between (II) and (III) can already be inferred from $H(k)$ being invariant under $(k,\theta)\rightarrow (-k,-\theta)$. The two nontrivial phases are reminiscent of the two dimerization phases, the so-called D1 and D2 phases, of the SSH model \cite{Atala2013}.

In one sense, both the Creutz and SSH models are (quasi-)one-dimensional models exhibiting nontrivial band topology and the existence of edge states. On the other hand, in the SSH model, there exist only two phases, both topologically non-trivial with $\varphi_{Zak}^{D1,D2}=\pm \pi/2$, separated by gap closing at a single quasimomentum at the Brillouin zone edge. In the Creutz model, it features three distinct phases, with the possibility of band crossings at either one or \textit{two} inequivalent quasimomenta. This leads to differences in the acquired Zak phase in the Creutz model. 

Moreover, as already noted by Creutz \cite{Creutz99}, the phase values $\theta=\pm\pi/2$ are special in that the spectrum becomes particle-hole symmetric. When $J_Y=0$, a complete basis takes the form of plaquette states (Fig. 1c) and flat band physics is also expected to be dominant here. 

To further characterize the physical distinction among the Creutz phases (besides the existence of edge states), a direct consequence which follows is that the Zak-phase-difference $\delta\varphi_{Zak}^{A-B}=\varphi_{Zak}^{A}-\varphi_{Zak}^{B}$ are
\beq
\delta\varphi_{Zak}^{\tr{II/III}-\tr{I}}=\pm\pi\tr{\ \ and\ \ }\delta\varphi_{Zak}^{\tr{II}-\tr{III}}=2\pi.
\eeq
These are experimentally measurable quantities \cite{King93, Atala2013,foot2}.

\textit{Case $\phi\neq 0$:} A nonzero $\phi$ diminishes the band crossing, except for $\phi=\pi$, Fig. 2b. 
The physical motivation for introducing $\phi$ as an additional phase follows from inspecting the topological part of the pseudospin Hamiltonian $H_{topo}(k)=\bm{h}(k)\cdot\bm{\sigma}$. With $\phi=0$, the band eigenstates, as a function of $k$, are determined by $(h_x, 0, h_z)$, are thus strictly planar, see right panel of Fig. 2a. The quantized Zak phase jumps discontinuously across the phase boundary. To establish an adiabatic connection between any two distinct phases (more precisely, to relate the two phases without encountering band crossing), we require a non-vanishing third pseudospin component in the Bloch Hamiltonian.  
With a broken inversion symmetry, it merely renormalizes $h_z$ while keeping $h_y=0$. We therefore introduce a phase imprint $\phi$ in the hopping along the rung direction giving $h_y\neq 0$. Crucially, due to the cross-link connectivity, this extra phase cannot be absorbed into a redefinition of $\theta$ but acts as an independent phase degree-of-freedom \cite{foot3}. The resulting model contains two gauge-invariant fluxes $(\alpha, \beta)$, both defined modulo $2 \pi$, and thereby giving way to an extended multiply-connected parameter space (Fig. 2b). We emphasize that with the network of line degeneracies it shows a topological scenario distinct from the SSH/RM. 

\begin{figure}
\begin{center}
\includegraphics[width=8.6cm]{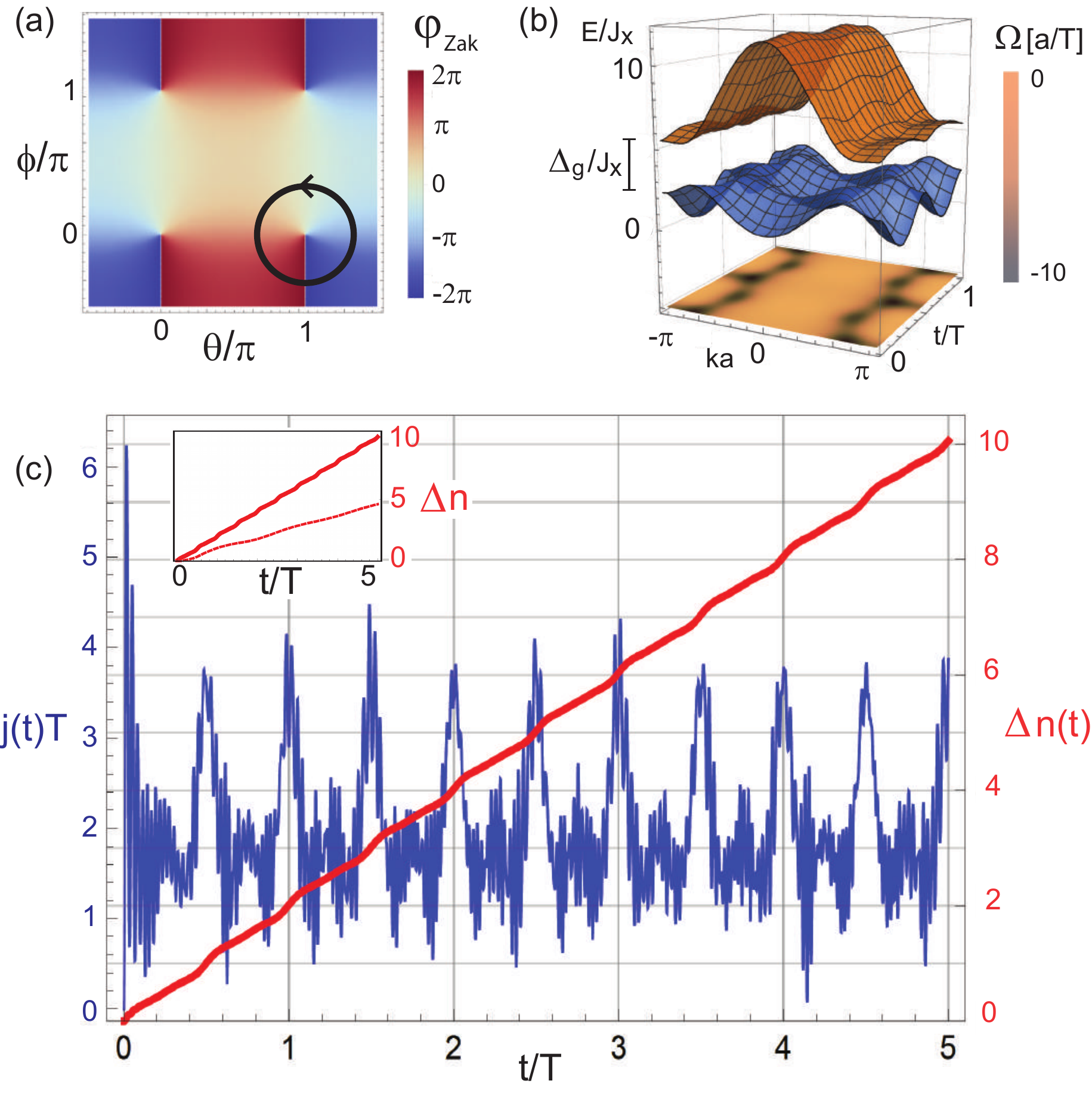}
\end{center}
\caption{(a) Zak phase as a function of $\theta$ and $\phi$ where the line of discontinuity (white line) is dependent on the choice of the eigenstate. The directed circle correspond to the $\gamma_1$ trajectory. (b) Evolution of the energy spectrum over a cycle. The density plot at the bottom is the Berry curvature associated with the lower energy band. (c) Time evolution of the current density $j(t)$ (blue line). The pumped charge up to time $t$ (red line). Inset: Adiabatic versus nonadiabatic charge transport (dash line). Chosen parameters: $J_Y/J_X=2$, $J_D/J_X=1.8$, $\hb\omega/J_X=0.1 \tr{ (adiabatic)}, 6.0 \tr{ (non-adiabatic)}$. }
\end{figure}

\paragraph{Quantum pumps and physical observables.---}
Following on the Introduction, two elementary pumping schemes are associated with closed trajectories enclosing the two kinds of boundary lines. They are trajectories connecting (II)-(III) phases and (I)-(II)/(III) phases, labeled, respectively, as $\gamma_{1,2}$ in Fig. 2c. The $\gamma_1$ pump turns out to be a novel kind, and in the following we will focus on its properties. At the end of this section, we comment on the other pumps. 

First, the $\gamma_1$ trajectory lies on the $(\theta,\phi)$-plane in the topological regime $J_D>J_D^c$, with all hopping strength $J_{D,X,Y}$ held fixed. On this plane, $\gamma_1$ encircles one of the band crossing points given by $(0,0)$ or $(\pi,0)$. For simplicity, we parametrize $\gamma_1$ as $\theta(t)=\cos(\omega t)+\pi$, $\phi(t)=\sin(\omega t)$, encircling $(\pi,0)$ with a time period $T$ for the Hamiltonian to return itself, i.e., $H(t+T)=H(t)$, $\omega=2\pi/T$ is the driving frequency (Fig. 1d). The chosen parameters $J_{X,Y,D}$ are such that the instantaneous energy band is well separated between the upper and lower branches during the time evolution. The energy gap $\Delta_g$ together with the driving frequency constitute an adiabatic condition, a condition easily met as we will show. The band topology characterizations are then given in terms of the Zak phase $\varphi_{Zak}(t)$, the Berry curvature $\Omega(k,t)$ and the Chern number $\mathcal{C}$ associated with the evolving band. 

From Fig. 3a, we see that $\gamma_1$ results in an absolute change of $4\pi$ in the Zak phase, twice that of $\delta\varphi_{Zak}^{II-III}$. An understanding can be drawn from the modern electric polarization theory by relating a $2\pi$ Zak-phase-difference as unit shift (in lattice constant unit) in the center-of-mass position of the Wannier states \cite{King93, Xiao10}. In this way, the $4\pi$ difference is related to twice the translation between the representative C1 and C2 coverings (Fig. 1c), a $2a$ shift, around which $\gamma_1$ traverses over one cycle. On the other hand, the Berry curvature $\Omega(k,t)$ associated with the evolution of the 1D energy band following $\gamma_1$ can be computed showing a definite sign throughout the 1BZ, see Fig. 3b. When integrated over one full cycle for each momentum state it gives the Chern number $|\mathcal{C}|=2$, consistent with the integrated Zak phase value.

On transport properties, we consider spinless fermions with the lower band fully filled \cite{foot5}. The time-evolution of the current density $j(t)\equiv (1/2\pi)\int_{1BZ}dk\,v_k(t)$ can be computed in a numerically exact way, following the full evolution of states according to the time-dependent Schr\"{o}dinger equation for the two-band model. Here $v_k(t)=\langle \psi_k(t)|\partial H(k,t)/\partial(\hbar k)| \psi_k (t)\rangle$ is the expectation value of the instantaneous velocity operator \cite{Xiao10} with the initial conditions $|\psi_k (0)\rangle=|u_k(0)\rangle $, where $k$ remains a good quantum number. Results for small driving $\hbar \omega/J_X=0.1$ are shown in Fig. 3c over several cycles (blue curve), where $\Delta_g=\mathcal{O}(J_X)$. Then, the pumped charge is obtained by integrating the current density $\Delta n(t)=\int^t_0 dt' \,j(t')$ (red curve) \cite{Xiao10}. In agreement with the adiabatic transport theory, the pumped charge over a full cycle in the limit of small driving gives the Chern number of the evolving band. In the inset we show the breakdown of adiabaticity with $\hbar \omega/J_X=6.0$, where $\Delta n(t)$ is no longer related to the band topology. We emphasize that topological charges only appear at multiple of complete cycles, while within a cycle the instantaneous charge pumping depends on the specific geometry of the trajectory. For example, at half cycle, it needs not, contrary to the one shown, be half the pumped charge of the full cycle. 

To sum up, the $\gamma_1$ quantum pump is remarkable because the effect of quantum interference (due to nontrivial lattice connectivity and fluxes) is only apparent in the structure of the complete basis, rather than in the Hamiltonian or the confining potential. It operates with time-varying fluxes only, without any `sliding' in the potential. 

Using the same characterization, the $\gamma_2$ trajectory can be realized in the $(\phi, J_D)$-plane  by encircling $(0,J_D^c)$ point with fixed $\theta\neq 0,\pi$. It results in the Zak phase difference of $2\pi$, the Chern number $|\mathcal{C}|=1$, and unit charge pumping. The pattern should now be clear: The topological charge pumping is related to the nature of the boundary line (i.e., the number of band crossing), its direction of encircling and in observing the adiabatic condition. As an illustration, we envisage a loop $\gamma_3$ encircling $(\theta,\phi)=(0,0)$ and $(0,\pi)$ in the topological regime, see Fig. 2c, resulting in a $|\mathcal{C}|=4$ quantum pump.

\begin{figure}
\begin{center}
\includegraphics[width=7.cm]{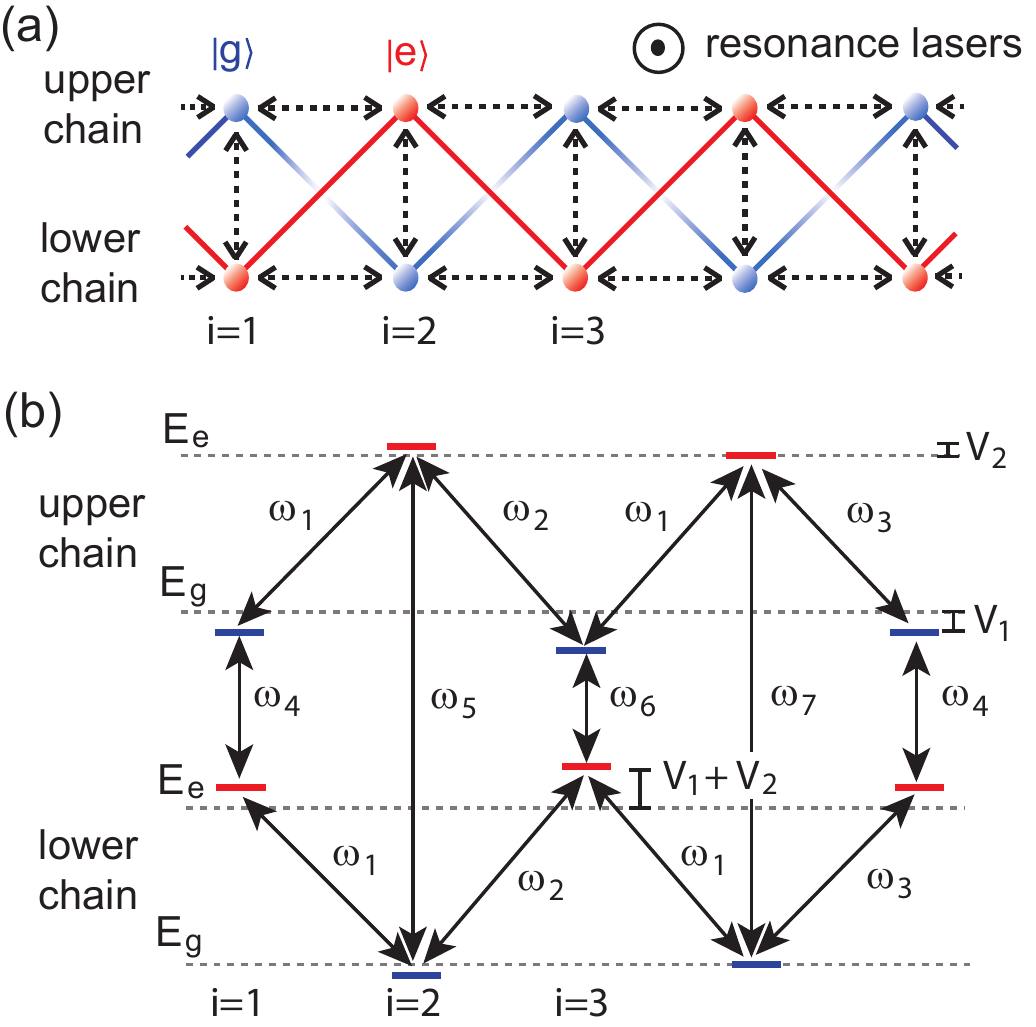}
\end{center}
\caption{(a) Proposed cold atomic setup with two-electron atoms trapped in a spatially shifted optical zigzag lattice \cite{Zhang15}. The inter-state conversions are induced by additional resonant lasers with tunable phases. (b) Spatially-dependent energy levels in a rectified optical superlattice (staggered in magnitude in the upper/lower chains) with two characteristic onsite energies $V_{1,2}$  \cite{Gerbier10}. Various laser frequencies are indicated with $\omega_1$-$\omega_7$.}
\end{figure}

{\it Cold atoms realization scheme\,-- }
We consider a quantum degenerate mixture of ground and long-lived excited states, denoted as $\{|g\rangle,|e\rangle\}$, realized with alkaline-earth atoms, suchs as ytterbium or strontium \cite{Livi16,Kolkowitz16}. 
They form a basis for the two-dimensional atom-laser coupling Hamiltonian \cite{Dalibard11}. As a first step, following standard interferometric means for realizing flexible lattice geometry \cite{Zhang15}, we consider trapping the two states separately in
two spatially shifted optical \textit{zigzag} lattices, see Fig. 4a. 
The second step follows the proposal of Ref. \cite{Gerbier10} by imposing an additional optical superlattice along the chain and shining resonant laser beams to induce $|g\rangle \leftrightarrow |e\rangle$ transitions. The superlattice lifts the spatial energy degeneracy on even-odd sites, see Fig. 4b, resulting in the requirement of seven resonance frequencies. By absorbing the appropriate photons, the laser-induced complex hoppings, $J_{X}e^{i\theta}$, $J_{Y}e^{i\phi}$ can be engineered in a site-resolved manner, see the Supplemental Material \cite{Suppl}.
We require the superlattice of Ref. \cite{Gerbier10} to be staggered in between the upper and lower chains, and a definite control of the phase locking among the laser beams. The final step involves setting the complex hoppings to be time-dependent on the adiabatic time scale, which can be achieved, for instance, by controlling the phase delay between the various lasers. As for physical observables, techniques to measure the shift in the atom cloud's center-of-mass, for example, have been developed for this kind of system \cite{Lohse16,Nakajima16,Lu16}. 

\paragraph{Conclusion and outlook.---}
We have studied the full topological features in the Creutz model as a distinct tight-binding model for realizing quantum charge pumping phenomenon. We applied the Peierl's phase substitution in the cross-link ladder giving way to an unique flux pattern with two gauge-invariant fluxes.
The elementary quantum pumps are associated with either one or two units pumped charge $|\mathcal{C}|=1,2$, a result of encircling the line degeneracy characterized by two kinds of band crossing. 
The $|\mathcal{C}|=2$ quantum pump emerges as a novel all-flux pump, suggesting a way to manipulate matter wave combining quantum interference effect and band topology. A cold atomic platform with tunable artificial gauge fields is shown to be a good candidate system for its realization. Serving as a microscopic chain to building higher dimensional systems, novel bulk and surface properties \cite{Hosur12} are expected to be uncovered.       

We thank Jean Dalibard for insightful comments on the cold atomic realization scheme. We thank Jean-No\"{e}l Fuchs, Lei Wang, Hui Zhai, Tin-Lun Ho, Mircea Trif and Gyu-Boong Jo  for helpful discussions. N. S. thank helpful discussions with Pengfei Zhang. This work was supported by Tsinghua University Initiative Research Programme and the Thousand Young Talents Program of China (L.-K. L.).

\widetext
\newpage
\section{Supplementary material}
\section{Laser scheme for the Creutz model}
\setcounter{equation}{0}
We give a description of the cold atomic proposal. Consider atoms with two internal long-lived states $\{|g\rangle,|e\rangle\}$ trapped in an optical zigzag lattices \cite{Zhang15a} which are spatially displaced horizontally by the lattice constant $a$ between the $g$ and $e$ states. The spatial overlap between nearest neighbour orbital wavefunctions (within the single band approximation) gives rise to the following tight-binding model:
\beq
H_{zz}=-J_D \sum_{m\in \mathbb{Z}} \biggl( |m+1,1\rangle_g\, _g\langle m,2| +|m+1,2\rangle_e\, _e\langle m,1|+\tr{h.c.} \biggr),\nn
\eeq
where $|m,1\rangle_g$ denotes an atom in the internal $g$ state at the position $m$ in the lower chain (e.g., $|m,2\rangle_e$ denotes internal state $e$, position $m$ and in the upper chain.) and $J_D$ is the hopping strength set by the optical lattice depth, see Fig.  5a. In addition, we impose a superlattice potential \cite{Gerbier10a} to modify spatially the onsite energies of $\{|g\rangle,|e\rangle\}$, with three characteristic energy parameters $V_1,V_2,V_1+V_2$ and they are staggered in between the upper and lower chains, resulting in the energy level diagram shown in the main text Fig. 4b and Fig. 5b. The various energy differences are given by 
\beq
&&\omega_1=E_0+V_1+V_2,\tr{\ \ }\omega_2=E_0+V_1+2 V_2,\tr{\ \ }\omega_3=E_0+V_1,\tr{\ \ }\nn\\
&&\omega_4=E_0+2 V_1,\tr{\ \ }\omega_5=E_0+2 V_2,\tr{\ \ }\omega_6=E_0+2 V_1+2 V_2,\tr{\ \ }\omega_7=E_0,
\eeq
and denoting the eight energy levels as $\{|g_i\rangle,|e_i\rangle\}$ for $i=1-4$. By turning on laser couplings $|g\rangle \leftrightarrow |e\rangle$ (with the Rabi frequency set to unity, for simplicity in notation), the atom-light coupling Hamiltonian \cite{Dalibard11a}, including the onsite energy shifts, in the rotating-wave-approximation is 
\beq
H_{a-l}&=&e^{-i\omega_1 t-i\phi_1}|e_1\rangle\, \langle g_1|+e^{-i\omega_2 t-i\phi_2}|e_1\rangle\, \langle g_2|+e^{-i\omega_1 t-i\phi_1}|e_2\rangle\, \langle g_2|+e^{-i\omega_3 t-i\phi_3}|e_2\rangle\, \langle g_1 |
\nn\\
&+&e^{-i\omega_1 t-i\phi_1}|e_3\rangle\, \langle g_3|+e^{-i\omega_2 t-i\phi_2}|e_4 \rangle\, \langle g_3|
+e^{-i\omega_1 t-i\phi_1}|e_4\rangle\, \langle g_4|+e^{-i\omega_3 t-i\phi_3}|e_3\rangle\, \langle g_4|\nn\\
&+&e^{-i\omega_4 t-i\phi_4}|e_3\rangle\, \langle g_1|+e^{-i\omega_5 t-i\phi_5}|e_1\rangle\, \langle g_3|+e^{-i\omega_6 t-i\phi_6}|e_4\rangle\, \langle g_2|+e^{-i\omega_7 t-i\phi_7}|e_2\rangle\, \langle g_4|+\tr{h.c.}\nn\\
&+&(-V_1) |g_1\rangle\, \langle g_1|+(E_0+V_2) |e_1\rangle\, \langle e_1|+(-V_1-V_2) |g_2\rangle\, \langle g_2|+(E_0) |e_2\rangle\, \langle e_2|\nn\\
&+&(E_0+V_1) |e_3\rangle\, \langle e_3|+(-V_2) |g_3\rangle\, \langle g_3|+(V_1+V_2) |e_4\rangle\, \langle e_4|+(0) |g_4\rangle\, \langle g_4|.
\eeq
\begin{figure}[!b]
\begin{center}
\includegraphics[width=15cm]{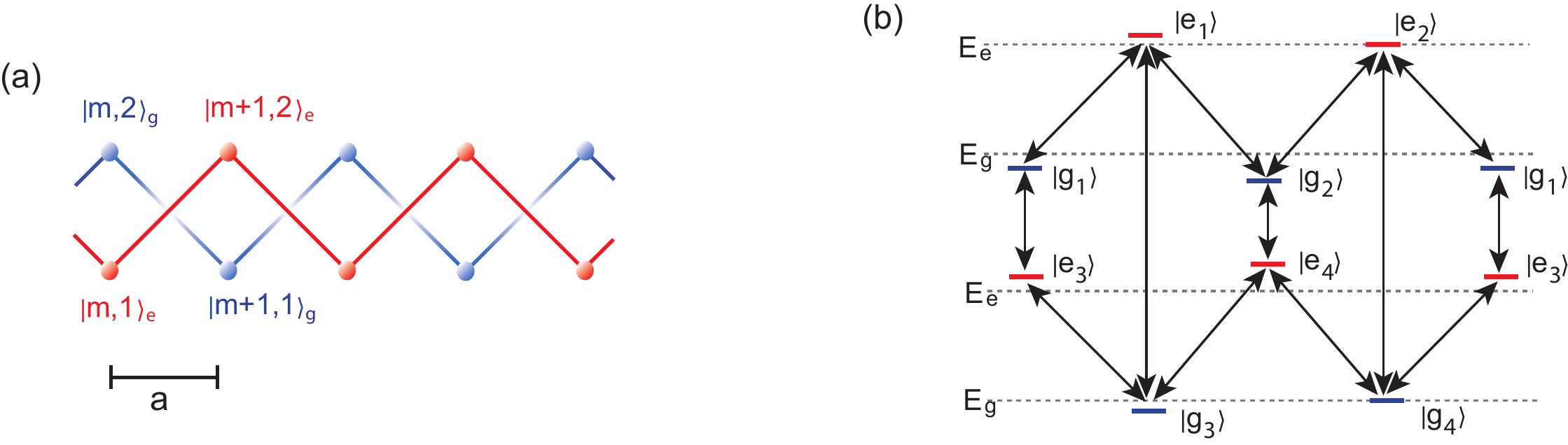}
\end{center}
\caption{(a) Two zigzag chains trapping $g$ and $e$ states, respectively. (b) The energy level diagram in the presence of the superlattice potential.}
\end{figure}
The phases $\phi_j$ for $j=1-7$ are associated with the initial seven resonance lasers. We now perform the following unitary transformations on the internal states:
\beq
&&|g_1\rangle \rightarrow  e^{-i V_1 t}|g_1\rangle;\tr{\ \ } |g_2\rangle \rightarrow  e^{-i (V_1+V_2) t}|g_2\rangle;\tr{\ \ }
|e_1\rangle \rightarrow  e^{i (E_0+V_2) t}|e_1\rangle;\tr{\ \ } |e_2\rangle \rightarrow  e^{i E_0 t}|e_2\rangle;\nn\\
&&|g_3\rangle \rightarrow  e^{-i V_2 t}|g_3\rangle;\tr{\ \ } |g_4\rangle \rightarrow  e^{-i 0 t}|g_4\rangle;\tr{\ \ \ \ \ \ \ \ \ \ }
|e_3\rangle \rightarrow  e^{i (E_0+V_1) t}|e_3\rangle;\tr{\ \ } |e_4\rangle \rightarrow  e^{i (E_0+V_1+V_2) t}|e_4\rangle;\nn
\eeq
and arrive at a time-independent atom-light coupling Hamiltonian 
\beq
H_{a-l}&=&e^{-i\phi_1}|e_1\rangle\, \langle g_1|+e^{-i\phi_2}|e_1\rangle\, \langle g_2|+e^{-i\phi_1}|e_2\rangle\, \langle g_2|+e^{-i\phi_3}|e_2\rangle\, \langle g_1 |
\nn\\
&+&e^{-i\phi_1}|e_3\rangle\, \langle g_3|+e^{-i\phi_2}|e_4 \rangle\, \langle g_3|
+e^{-i\phi_1}|e_4\rangle\, \langle g_4|+e^{-i\phi_3}|e_3\rangle\, \langle g_4|\nn\\
&+&e^{-i\phi_4}|e_3\rangle\, \langle g_1|+e^{-i\phi_5}|e_1\rangle\, \langle g_3|+e^{-i\phi_6}|e_4\rangle\, \langle g_2|+e^{-i\phi_7}|e_2\rangle\, \langle g_4|+\tr{h.c.}\nn
\eeq
We demand the laser phases to be locked at $\phi_1=-\phi_2=-\phi_3=-\theta$, $\phi_4=-\phi_5=\phi_6=-\phi_7=-\phi$. To take advantage of the gauge freedom, we note that there are actually four observable gauge-invariant phases given
\beq
\Phi_1=2\phi_1-\phi_4-\phi_5,\tr{\ \ }\Phi_2=2\phi_2-\phi_5-\phi_6,\tr{\ \ }\Phi_3=2\phi_1-\phi_6-\phi_7,\tr{\ \ }\Phi_4=2\phi_3-\phi_4-\phi_7.
\eeq
The phase locking condition amounts to two further conditions on the gauge-invariant phases $\Phi_1+\Phi_2=\Phi_3+\Phi_4=0$. The latter can facilitate in the design of the coupling of the various laser phases. 
\\
\\
By controlling the Rabi frequencies and taking into account the finite spatial overlap between the single-band wavefunctions we finally obtain the Hamiltonian for the generalized Creutz scheme
\beq
H_{zz}+H_{a-l}&=&H_{zz}+ \hbar \Omega_1 \sum_{m\in \mathbb{Z}}  \biggl(e^{i \phi} |m,1\rangle_e\, _g\langle m,2 |+ \tr{h.c.}\biggr)\nn\\
&&+ \hbar \Omega_2 \sum_{m\in \mathbb{Z}} \biggl( e^{i\theta} |m+1,2\rangle_e \, _g\langle m,2 |+ e^{-i\theta} |m+1,1\rangle_g \, _e\langle m,1 |+\tr{h.c.}\biggr)\nn
\eeq


\begin{thebibliography}{99}
\bibitem{foot1} The energy spectrum possesses several elementary symmetries, namely that $E(k)=E(-k)$, $E(\theta)=E(-\theta)$, $E(\theta)=-E(-\pi+\theta)$ and $E(\phi)=E(-\phi)$.
\bibitem{foot5} The Fermi level at half-filling generally does not lie in the gap even when the two energy bands are isolated. However, increasing $J_D$ (while keeping all other parameters constant) can bring the system into a half-filled gapped state without undergoing a topological band transition.
\bibitem{foot2} We emphasize that the Zak phase captures the absolute winding (up to an overall sign) of the pseudospin, which is more than just a phase quantity being defined modulo $2\pi$.
\bibitem{foot3} Other gauge choice is also possible with, for example, the additional phase imprinted along the cross link hopping. However, it can be shown to be gauge-equivalent to the Hamiltonian (1).
\end{thebibliography}
\end{document}